\documentclass[runningheads]{llncs}

\usepackage[utf8]{inputenc} 
\DeclareUnicodeCharacter{202F}{\,}
\usepackage[T1]{fontenc}    
\usepackage{hyperref}       
\usepackage{url}            
\usepackage{booktabs}       
\usepackage{multirow}       
\usepackage{amsfonts}       
\usepackage{nicefrac}       
\usepackage{microtype}      
\usepackage{lipsum}
\usepackage{graphicx}
\graphicspath{ {./images/} }
\usepackage{pgf}
\usepackage{tikz}
\usepackage{pgfplots}
\pgfplotsset{compat=1.18}
\usepackage{subcaption}
\newcommand{\LLMs}{Large Language Models}
\newcommand{\SMNGPtwo}{SuperMUC-NG Phase 2}
\newcommand{\device}{Intel Data Center GPU Max~1550}
\newcommand{\deephyper}{\texttt{DeepHyper}}

\title{A Scalable Recipe on \SMNGPtwo{}: Efficient Large-Scale Training of Language Models}
\titlerunning{Efficient Large-Scale Training of LLMs on \SMNGPtwo{}}

\author{Ajay Navilarekal Rajgopal\inst{1} \and Nikolai Solmsdorf\inst{2}}
\institute{Leibniz Supercomputing Centre of the Bavarian Academy of Science and Humanities, Boltzmannstr. 1, 85748 Garching, Germany \\
\email{Ajay.Navilarekal@lrz.de}
\and
Intel Deutschland GmbH, Dornacher Straße 1, 85622 Feldkirchen, Germany\\
\email{nikolai.solmsdorf@intel.com}}

\begin{document}

\maketitle


\begin{abstract}
Large Language Models (LLMs) continue to demonstrate superior performance with increasing scale, yet training models with billions to trillions of parameters requires staggering computational resources—%
e.g., a one-trillion-parameter GPT-style model requires an estimated 120 million exaflops.
This challenge necessitates efficient distributed training strategies on cutting-edge High-Performance Computing (HPC) infrastructure.
In this work, we explore the \SMNGPtwo{} (SMNG-P2) system at the Leibniz Supercomputing Centre (LRZ) in Garching, Germany, equipped with \device{} accelerators to extract the necessary computational power. 
We enable and investigate a comprehensive \textbf{recipe} of parallel training techniques, including tensor parallelism, pipeline parallelism, and sharded data parallelism, essential for facilitating the training of \LLMs{} up to 175 billion-parameter scale on SMNG-P2.
Through empirical assessment and extensive hyperparameter tuning, we analyze the complex interplay among these techniques and determine their impact on GPU computational efficiency.
We identify an optimized combined strategy that yields high throughput and enables the efficient training of \LLMs{} of varying sizes.
Specifically, for the 175B model, we achieved per-tile throughput of 10\% of theoretical peak per-tile \texttt{bf16} FLOPs, employing an out-of-the-box publicly available software stack, utilizing standard distributions without further modification. This approach ensures broad accessibility, as our methodology can be replicated by any user on the \SMNGPtwo{} system without the need for porting or specialized software engineering.
Furthermore, we achieved 93\% weak scaling efficiency on 128 nodes alongside strong scaling efficiency of 82\% on 128 nodes.
This scalable recipe provides a crucial blueprint for efficiently utilizing advanced exascale systems for next-generation foundational model development.
\end{abstract}

\keywords{Large Language Models, Transformer Training, Distributed Training, Model Parallelism, Tensor Parallelism, Pipeline Parallelism, ZeRO, Megatron-DeepSpeed, SuperMUC-NG Phase 2, Intel Data Center GPU Max 1550}


\section{Introduction}
Large Language Models (LLMs) continue to improve with scale, but training at tens to hundreds of billions of parameters requires careful coordination of compute, memory, and communication on modern HPC systems.

In this work, we study SuperMUC-NG Phase~2 at the Leibniz Supercomputing Centre, a production accelerator system based on Intel Data Center GPU Max~1550.
\cite{supermuc_ng_phase2_lrz_2024,sng2_pilotphase_lrz_2024,intel_gpu_max_1550}
Our goal is a reproducible, out-of-the-box recipe for GPT-style training without custom kernels or hardware-specific modifications.
Using Megatron-DeepSpeed, we replicate the methodology of \cite{optimizing_frontier} and evaluate scaling behavior of a model with 175B parameters.
\cite{mt_nlg_530b,megatron_lm_2019,zero_optimizations}
We characterize tensor, pipeline, and sharded data parallelism, identify key cross-node bottlenecks, and provide practical configuration guidance for sustained throughput on SuperMUC-NG Phase~2 and similar production systems.

\paragraph{Outline}
Section~\ref{sec:parallelism} covers parallelism strategies, Section~\ref{sec:hardware} the system environment, Section~\ref{sec:benchmarking} benchmarking, Section~\ref{sec:hparam} automated tuning, Section~\ref{sec:scaling} scaling analysis, and Section~\ref{sec:conclusion} conclusions.

\section{Parallelism Strategies}
\label{sec:parallelism}
 \subsection{GPT Model}

GPT (Generative Pre-trained Transformer) models are decoder-only Transformers for autoregressive language modeling.\cite{transformer_attention,gpt_1,gpt_2,gpt_3}
For a model with hidden size $d$, $L$ layers, and vocabulary size $V$, a standard parameter estimate is
$P_{\mathrm{total}} \approx 12Ld^2 + Vd$, where the $12Ld^2$ term dominates at scale.

In this work we use three model sizes (3.6B, 20B, and 175B), with memory requirements summarized in Table~\ref{tab:model_memory}.
Under mixed precision, each parameter contributes 16 bytes in total across master weights, compute weights, gradients, and Adam states.

\begin{table}[]
\centering
\renewcommand{\arraystretch}{1.2}
\begin{tabular}{c|ccc}
\hline
\multirow{2}{*}{}     & \multicolumn{3}{c}{Memory} \\ \cline{2-4}
                      & 3.6B    & 20B    & 175B    \\ \hline
Parameters (6x)       & 21.6GB  & 120GB  & 1.05TB  \\ \hline
Gradients (2x)        & 7.2GB   & 40GB   & 0.35TB  \\ \hline
Optimizer States (8x) & 28.8GB  & 160GB  & 1.4TB   \\ \hline
\textbf{Total}        & \textbf{57.6GB}  & \textbf{320GB}  & \textbf{2.8TB}   \\ \hline
\end{tabular}
\caption{Memory requirements of the different model sizes}
\label{tab:model_memory}
\vspace{-27pt}
\end{table}

A single \SMNGPtwo{} node has 4 \device{} accelerators, each split into 2 tiles (8 tiles per node).
Each accelerator provides 128 GB HBM2e (64 GB per tile), so model parallelism is required to fit the 20B and 175B models.

 \subsection{Tensor Parallelism}
Tensor parallelism targets the largest matrix multiplications in transformer blocks by splitting weight matrices across devices and executing partial products in parallel.\cite{megatron_lm_2019}
In GPT-like models, this is most commonly applied to the projection layers in multi-head attention (the $QKV$ and output projections) and the two MLP projections.
Each accelerator owns a shard of the parameters and activations, computes its local slice, and then participates in collectives (typically \texttt{all-reduce} or \texttt{all-gather}) to assemble the full layer output.
This sharding keeps per-device memory within limits while preserving the layer-wise computation structure of the transformer.

The performance of tensor parallelism is strongly tied to communication costs because the collectives occur at every layer.

 \subsection{Pipeline Parallelism}
Pipeline parallelism partitions transformer layers into $PP$ stages and overlaps stage execution across micro-batches.\cite{pipedream,gpipe}
Its main overhead is the pipeline bubble, which shrinks as the micro-batch count $M$ increases.
In 1F1B scheduling, bubble overhead scales approximately as $PP/M$, and interleaving further reduces idle time by splitting stages into virtual sub-stages.\cite{megatron_lm}

 \subsection{Sharded Data Parallelism}
Sharded data parallelism (SDP) reduces memory use by partitioning model states across data-parallel ranks instead of full replication.
DeepSpeed implements this through ZeRO: stage~1 shards optimizer states, stage~2 shards optimizer states and gradients, and stage~3 also shards parameters.\cite{zero_optimizations}
Higher ZeRO stages improve memory efficiency but increase communication overhead.

 \subsection{Megatron-DeepSpeed}

No single parallelism dimension is sufficient at this scale; practical training combines tensor, pipeline, and data parallelism (3D parallelism) to balance memory, communication, and utilization.
Megatron-LM provides optimized tensor/pipeline primitives, and Megatron-DeepSpeed adds ZeRO-based sharding and runtime support for large-scale training.\cite{megatron_lm_2019,megatron_lm_github,mt_nlg_530b,megatron_deepspeed_github}

For this work we use Megatron-DeepSpeed \texttt{v2.4} out-of-the-box with DeepSpeed \texttt{0.16.9}, PyTorch \texttt{2.8.0}, and XPU support from Intel Extension for PyTorch \texttt{2.8.0}.\cite{intel_ipex_docs,intel_ipex_github}
We avoid hardware-specific kernels so the workflow remains reproducible and directly usable on \SMNGPtwo{} with public software stacks.

\section{Hardware Overview}
\label{sec:hardware}

Experiments run on SuperMUC-NG Phase~2 at LRZ, a production accelerator partition with 240 GPU-capable nodes (234 compute, 2 spare, 4 login).\cite{supermuc_ng_phase2_lrz_2024,sng2_pilotphase_lrz_2024}

\subsection{Compute and Accelerator Configuration}

Each compute node has two 4th~Gen Intel Xeon Scalable CPUs (112 cores/node), 512\,GB DDR5 memory, and four Intel Data Center GPU Max~1550 accelerators.\cite{intel_gpu_max_1550}
The GPU Max~1550 implements a multi-tile architecture with 128\,GB HBM2e per accelerator and Intel Xe~Link for intra-node GPU communication.\cite{intel_gpu_max_1550}
Across the machine, this corresponds to 960 GPUs and approximately 123\,TB of aggregate GPU memory.\cite{supermuc_ng_intro_v3}

\subsection{Interconnect and Storage}

Nodes are connected via a fat-tree NVIDIA/Mellanox HDR InfiniBand fabric with two HDR interfaces per node (400\,Gbit/s aggregate injection bandwidth).\cite{supermuc_ng_intro_v3}
Storage uses a DAOS tier (about 1\,PB usable, over 750\,GB/s aggregate write bandwidth), with transparent access to Phase~1 parallel file systems (HPPFS and DSS).\cite{supermuc_ng_intro_v3}

\subsection{Operational Conditions Relevant to Scaling}

In production, GPU power is capped to 450\,W per accelerator (nominal 600\,W).\cite{power_capped_lrz_2025}
Therefore, reported performance reflects sustained production operation rather than unconstrained peak settings.
All experiments use the standard LRZ software environment (SLES + SLURM) and out-of-the-box software stacks without custom kernels or framework modifications.

\section{Benchmarking Parallelism Strategies}
\label{sec:benchmarking}

In this section, we present experiments on different parallelism strategies. All benchmarks use \texttt{bf16} precision. To isolate system-level performance effects, each configuration is run for 10 training steps rather than full convergence; in our measurements, throughput stabilizes within this warmup horizon.

\subsection{Tensor Parallelism}
\paragraph{Experimental setup.}
We use a 3.6B-parameter model to study how the tensor-parallel (TP) degree impacts sustained throughput.
To isolate the effect of TP, we fix the pipeline-parallel degree to $1$ and keep the micro-batch size as well as the number of micro-batches per optimizer step constant across all runs.

\paragraph{Sweep configuration.}
We sweep TP over $\{4,8,16\}$ tiles and increase the world size proportionally so that the global batch size remains constant.
This ensures that observed performance differences primarily reflect changes in communication volume and placement (intra-node vs.\ inter-node), rather than changes in statistical efficiency or optimizer behavior.

\begin{figure}[ht]
    \centering
    \begin{tikzpicture}
    \begin{axis}[
        ybar,
        width=10cm,
        height=7cm,
        xlabel={Tensor Parallel Degree},
        ylabel={Throughput (TFLOPs)},
        symbolic x coords={4, 8, 16},
        xtick=data,
        ymin=0,
        enlarge x limits=0.2,
        grid=major,
        grid style={dashed, gray!30}
    ]
        \addplot [fill=blue!40, draw=blue!70]
            table [x={Tensor Parallel Degree}, y=Throughput, col sep=comma] {tp/tp_results.csv};
    \end{axis}
\end{tikzpicture} 
    \caption{Throughput vs TP degree for 3.6B model}
    \label{fig:tp_sweep}
    \vspace{-20pt} 
\end{figure}
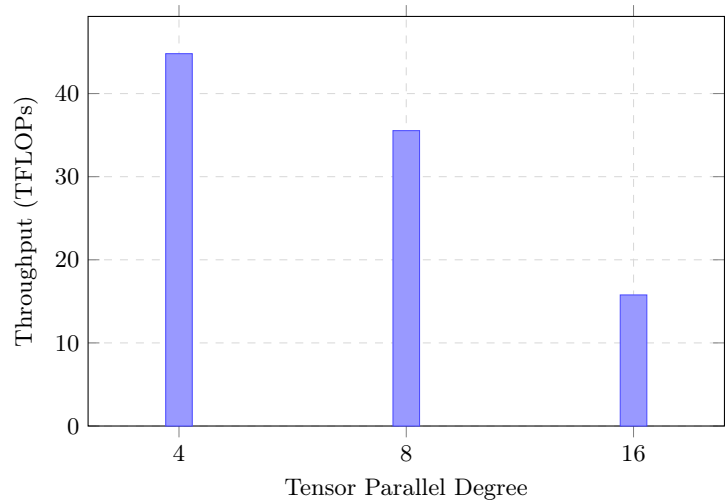

\paragraph{Observed throughput trend.}
As shown in Figure~\ref{fig:tp_sweep}, throughput drops sharply once the TP degree crosses the node boundary (i.e., $\mathrm{TP} > 8$ on our system).
For $\mathrm{TP}\le 8$, collective operations are confined to a single node and benefit from high-bandwidth intra-node GPU connectivity.
When $\mathrm{TP}=16$, the same collectives span beyond a single node and become sensitive to the lower bandwidth and higher latency of the interconnect.

\paragraph{Communication bottleneck explanation.}
Tensor parallelism introduces frequent synchronization points, dominated by \texttt{all-reduce} operations on activation and gradient tensors.
Beyond a single node, the increased cost of these high-frequency collectives limits overlap with compute and reduces overall pipeline utilization, resulting in the observed throughput degradation.

\paragraph{Conclusion.}
Based on these results, we limit tensor parallelism to a single node.
In our setup, this corresponds to choosing $\mathrm{TP}\le 8$ for subsequent experiments, and scaling to larger world sizes via data parallelism and/or pipeline parallelism instead.

\subsection{Pipeline Parallelism}
\paragraph{Experimental setup.}
We use a 20B-parameter model to study how pipeline-parallel degree affects sustained throughput.
The primary factor is the pipeline bubble, whose size scales with the ratio $\frac{PP}{M}$, where $PP$ is the number of pipeline stages and $M$ is the number of micro-batches per optimizer step.

\paragraph{Sweep configuration.}
We run three targeted experiments to isolate the effect of this ratio:
(i) keep $PP$ constant while varying $M$,
(ii) keep $M$ constant while varying $PP$, and
(iii) keep $PP/M$ constant by scaling $PP$ and $M$ together.

\subsubsection{Experiment 1}
\begin{figure}[ht]
    \centering
    \begin{tikzpicture}
    \pgfplotsset{set layers}
    
    \begin{axis}[
        name=mainplot,
        width=\linewidth,
        height=7cm,
        axis y line*=left,
        grid=major,
        xmin=0, xmax=520, 
        ymin=0, ymax=55, 
        color=black,
        xlabel={}, ylabel={}
    ]
        \addplot[mark=x, black] table [x=M, y=Throughput, col sep=comma] {pp/exp1.csv};
        \label{plot:throughput}
    \end{axis}

    \begin{axis}[
        width=\linewidth,
        height=7cm,
        axis y line*=right,
        axis x line=none,
        xmin=0, xmax=520, 
        ymin=0, ymax=4, 
        color=black,
        ylabel={},
        legend style={at={(0.97,0.03)}, anchor=south east} 
    ]
        \addplot[mark=o, dashed, black] table [x=M, y=Gain, col sep=comma] {pp/exp1.csv};
        \label{plot:gain}
        
        \addlegendimage{/pgfplots/refstyle=plot:throughput}\addlegendentry{Throughput}
        \addlegendimage{/pgfplots/refstyle=plot:gain}\addlegendentry{Gain}
    \end{axis}

    \node[below=0.8cm] at (mainplot.south) {\textbf{Number of Micro-batches}};
    \node[rotate=90, above, xshift=0.1cm, yshift=0.7cm] at (mainplot.west) {\textbf{Throughput (TFLOPs)}};
    
    \node[rotate=-90, below, xshift=0.1cm, yshift=0.9cm] at (mainplot.east) {\textbf{Gain}};

\end{tikzpicture}
    \caption{Experimental results showing Throughput and Gain vs. Number of Micro-batches.}
    \label{fig:experiment1-pp}
    \vspace{-27pt} 
\end{figure}
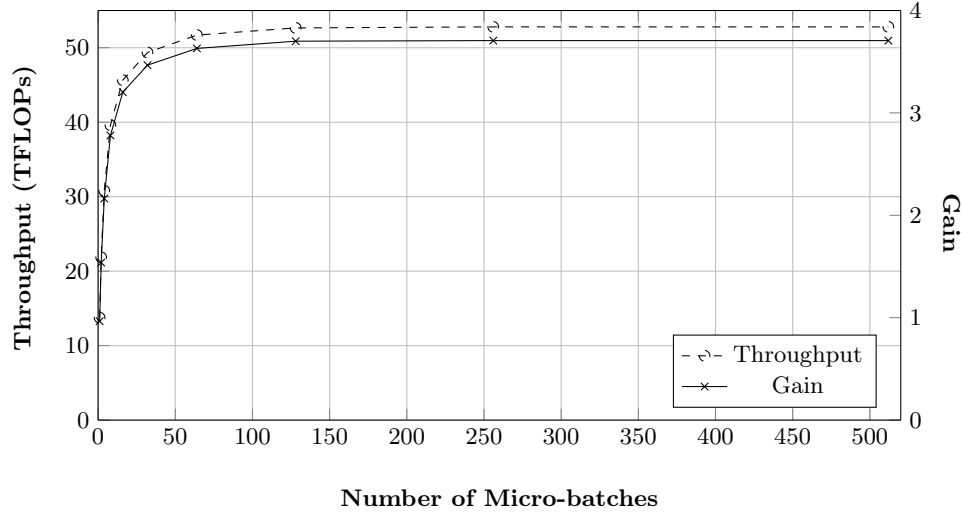

We keep the number of pipeline stages $PP$ fixed and vary the number of micro-batches $M$.
Throughput increases initially as the bubble is amortized, then plateaus beyond a certain $M$, yielding diminishing returns (see Figure~\ref{fig:experiment1-pp}).
Larger $M$ also implies a larger global batch size per model replica, which can constrain the attainable degree of data parallelism.

\subsubsection{Experiment 2}
We keep the number of micro-batches $M$ fixed and increase $PP$.
As expected, increasing $PP$ at fixed $M$ grows the bubble size and reduces throughput.
This trend is observed in the measured results (see Figure~\ref{fig:experiment2_3-pp}).

\subsubsection{Experiment 3}
\begin{figure}[ht]
    \centering
    \begin{tikzpicture}
    \begin{axis}[
        width=\linewidth,
        height=7cm,
        ybar,                
        bar width=10pt,      
        xlabel={PP},
        ylabel={Throughput (TFLOPs)},
        symbolic x coords={4, 8, 16}, 
        xtick=data,
        ymin=0,
        enlarge x limits=0.15,
        legend style={at={(0.5,-0.2)}, anchor=north, legend columns=-1},
        grid=major,
        grid style={dashed, gray!30}
    ]
        \addplot[fill=blue!60, draw=black] 
            table [x=PP, y=Throughput, col sep=comma] {pp/exp2.csv};
        \addlegendentry{Constant M}

        \addplot[fill=orange!70, draw=black] 
            table [x=PP, y=Throughput, col sep=comma] {pp/exp3.csv};
        \addlegendentry{Constant $\frac{PP}{M}$}
        
    \end{axis}
\end{tikzpicture}
    \caption{Experimental results showing Throughput vs. Number of pipeline stages}
    \label{fig:experiment2_3-pp}
\end{figure}
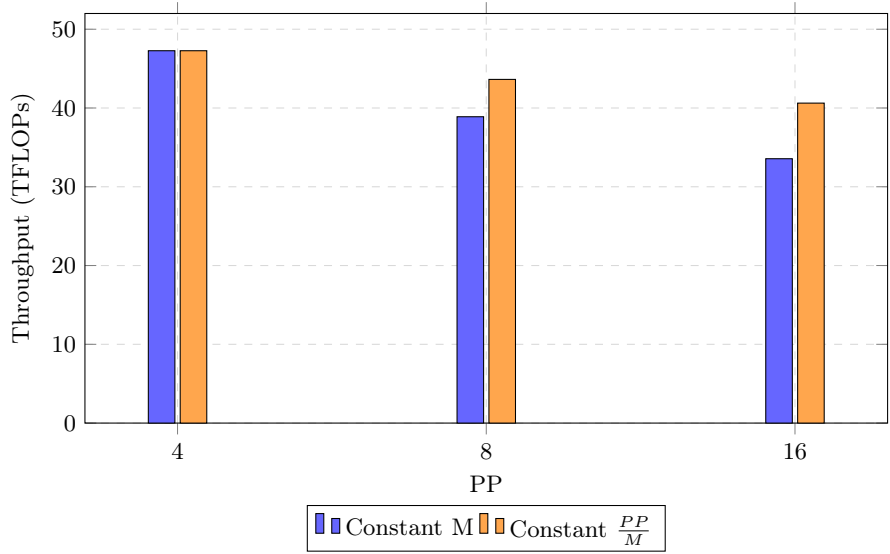

To offset the performance loss from larger $PP$, we increase $M$ simultaneously to keep $PP/M$ constant.
Under this constraint, throughput remains stable, which is confirmed by the results (see Figure~\ref{fig:experiment2_3-pp}).

\paragraph{Conclusion.}
Throughput is primarily governed by the $PP/M$ ratio: increasing $M$ reduces the bubble and improves utilization, while increasing $PP$ at fixed $M$ degrades throughput.
Keeping $PP/M$ approximately constant maintains stable performance, but larger $M$ implies larger effective batch sizes and limits data-parallel scaling.
Based on these results, we choose a moderate pipeline depth and sufficient micro-batching to balance utilization with batch-size constraints in subsequent runs.

\section{Automated Parallelism Strategy Search}
\label{sec:hparam}
Owing to the complex interplay of the different training parameters, we automate the search by treating the determination of these parameters as a Bayesian Optimization problem.

\paragraph{Problem setup.}
We frame the selection of distributed training parameters as a black-box optimization problem: for a given model configuration, we seek the parallelism setting that maximizes sustained throughput under memory constraints.
We measure throughput as effective model TFLOPs/s, computed from the forward and backward passes and reported by Megatron-DeepSpeed.
This metric is appropriate because it directly captures end-to-end training efficiency for a fixed model, integrating both compute utilization and communication overhead, and it is comparable across parallelism configurations under the same workload assumptions.
To isolate the effect of model parallelism choices, we fix the data-parallel size to $1$ and assume that the search is not constrained by the number of available nodes (i.e., we evaluate candidate configurations with sufficient resources to run).
As in the benchmarking section, all runs use \texttt{bf16} precision and execute 10 training steps per configuration to obtain stable throughput estimates.

\paragraph{Optimization method.}
We use Bayesian Optimization (BO) to search the mixed discrete/continuous space with fewer expensive trials than grid or random search.
BO uses a surrogate model and acquisition function to prioritize informative configurations, which is critical when each evaluation is a multi-node training run.
The BO loop is implemented with \deephyper{} \texttt{0.8.1}\cite{deephyper_joss_2025}, which supports asynchronous evaluations on single- and multi-node HPC systems.

\paragraph{Search space.}
Each candidate configuration is defined by the following hyperparameters:
\begin{itemize}
\item pipeline-parallel degree (PP),
\item tensor-parallel degree (TP),
\item micro-batch size (MBS),
\item gradient accumulation steps (GAS), i.e., the number of micro-batches accumulated before each optimizer step.
\end{itemize}
We additionally impose a fixed budget on the total number of evaluations to keep the tuning cost manageable.
This is a practical requirement on typical HPC systems, where large jobs must fit within finite allocation budgets, queue-time constraints, and per-job wall-time limits.
By capping the number of BO trials, the workflow remains feasible for production environments while still enabling systematic exploration of the most influential parallelism settings.

\paragraph{Execution workflow.}
We run BO on the 175B model as a representative large-scale case (Table~\ref{tab:hpo_space}).
At each iteration, \deephyper{} proposes $(\mathrm{PP},\mathrm{TP},\mathrm{MBS},\mathrm{GAS})$, we launch the corresponding SLURM job via \texttt{sbatch}, and we return parsed throughput to the optimizer.
Failed runs (e.g., OOM, invalid factorization, timeout) receive a penalized \texttt{F}-value so BO learns to avoid infeasible regions.

\begin{table}[t]
  \caption{Hyperparameter search space and results for automated tuning of the 175B model.}
  \label{tab:hpo_space}
  \centering
  \setlength{\tabcolsep}{10pt}
  \begin{tabular}{c|c|c}
    \toprule
    Hyperparameter & Range & Value \\
    \midrule
    Pipeline parallelism (PP) & $\{12,16,20,24\}$ & $16$ \\
    Tensor parallelism (TP) & $\{4,8\}$ & $8$ \\
    Micro-batch size (MBS) & $[1,10]$ & $3$ \\
    Gradient accumulation steps (GAS) & $\{25,50,100\}$ & $100$ \\
    \bottomrule
  \end{tabular}
\end{table}

\begin{figure}[ht]
    \centering
    \begin{tikzpicture}
    \begin{axis}[
        width=\textwidth, 
        height=7cm,
        xlabel={Evaluation},
        ylabel={Achieved GPU Throughput},
        legend pos=north west,
        grid=major,
    ]
        \addplot[only marks, mark=*, green] 
            table [x=job_id, y=objective, col sep=comma] {hparamSearch/success_data.csv};
        \addlegendentry{Success}

        \addplot[only marks, mark=x, red, mark size=3pt] 
            table [x=job_id, y=objective, col sep=comma] {hparamSearch/error_data.csv};
        \addlegendentry{Failure}
    \end{axis}
\end{tikzpicture} 
    \caption{Search Trajectory}
    \label{fig:bo_trajectory}
\end{figure}
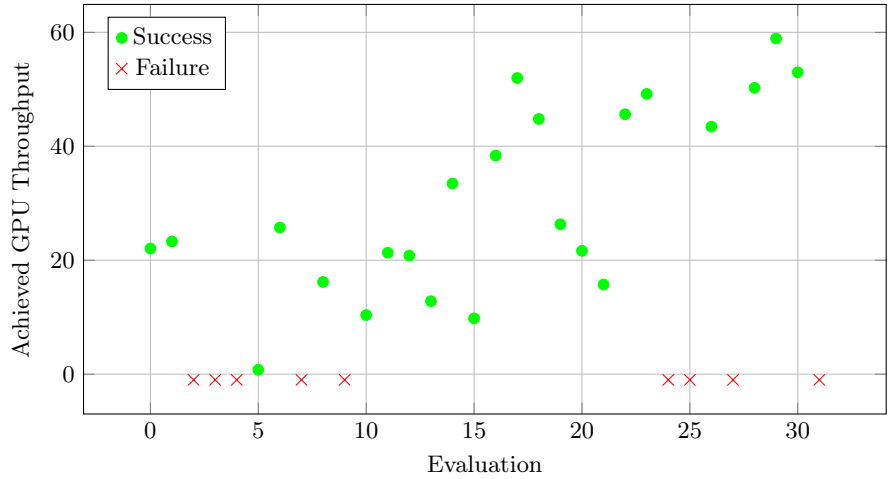

\paragraph{Observed behavior.}
Figure~\ref{fig:bo_trajectory} shows best-so-far TFLOPs/s versus evaluation index.
The trajectory improves from low-performing initial points toward higher-throughput configurations, while red markers indicate failures (mostly OOM).
The best observed throughput is 57 TFLOPs/s per-tile ($\sim10\%$ of theoretical peak \texttt{bf16} FLOPs per-tile), consistent with utilization levels reported in prior large-scale studies under communication constraints.\cite{optimizing_frontier,palm_scaling}
We use the best configuration from Table~\ref{tab:hpo_space} for the scaling experiments.

%

\section{Scaling Performance}
\label{sec:scaling}

Scaling \LLMs{} requires sustaining per-tile utilization as tile count increases. We use sharded data parallelism (ZeRO-1)\cite{zero_optimizations} and scale to 128 nodes on \SMNGPtwo{} (1024 tiles, over 50\% of the system's compute nodes) to quantify weak and strong scaling behavior.

\subsection{Weak Scaling}
In weak scaling, we increase global batch size with tile count while keeping per-tile workload approximately constant. Scaling to 128 nodes (8$\times$ baseline) yields weak-scaling efficiency of $\sim93\%$, indicating near-ideal throughput preservation over the evaluated range.

\subsection{Strong Scaling}
In strong scaling, we keep global batch size fixed while increasing tile count. At 128 nodes (8$\times$ baseline), strong-scaling efficiency is $\sim82\%$, lower than weak scaling because reduced work per replica increases synchronization and communication overhead.

\begin{figure}[!t]
    \centering
    \begin{subfigure}{0.49\textwidth}
        \centering
        \begin{tikzpicture}
    \begin{axis}[
        width=0.95\linewidth, 
        height=3.8cm,       
        xlabel={Number of tiles},
        ylabel={Weak Scaling Efficiency \%},
        ymin=0, ymax=105,      
        ytick distance=20,     
        xmin=0, xmax=1100,      
        xtick distance=200,     
        tick label style={font=\scriptsize},
        label style={font=\small},
        grid=major,
        color=black       
    ]
        \addplot[mark=x, black] 
            table [x=Number of tiles, y=Weak Scaling Efficiency, col sep=comma] {scaling/scaling_efficiency.csv};

    \end{axis}
\end{tikzpicture}
        \caption{Weak scaling efficiency}
        \label{fig:weak_scaling}
    \end{subfigure}\hfill
    \begin{subfigure}{0.49\textwidth}
        \centering
        \begin{tikzpicture}
    \begin{axis}[
        width=0.95\linewidth, 
        height=3.8cm,       
        xlabel={Number of tiles},
        ylabel={Strong Scaling Efficiency \%},
        ymin=0, ymax=105,      
        ytick distance=20,     
        xmin=0, xmax=1100,      
        xtick distance=200,     
        tick label style={font=\scriptsize},
        label style={font=\small},
        grid=major,
        color=black       
    ]
        \addplot[mark=x, black] 
            table [x=Number of tiles, y=Strong Scaling Efficiency, col sep=comma] {scaling/scaling_efficiency.csv};

    \end{axis}
\end{tikzpicture}
        \caption{Strong scaling efficiency}
        \label{fig:strong_scaling}
    \end{subfigure}
    \caption{Weak and strong scaling efficiencies}
    \label{fig:scaling}
\end{figure}
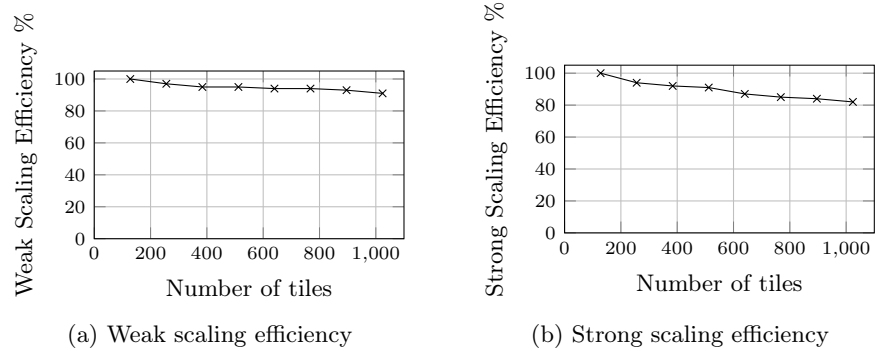

Figure~\ref{fig:scaling} summarizes both results: weak scaling remains near-ideal, while strong scaling degrades gradually at larger scales due to communication costs. This trend is consistent with \cite{optimizing_frontier}.

\section{Conclusion}
\label{sec:conclusion}
We demonstrate large-scale training on \SMNGPtwo{} by balancing tensor, pipeline, and data parallelism across hundreds of \device{} accelerators.
\cite{supermuc_ng_phase2_lrz_2024,sng2_pilotphase_lrz_2024}
Our results reproduce the methodology of \cite{optimizing_frontier} on an alternative production platform and confirm competitive sustained throughput at scale.

\paragraph{Checklist}
\begin{itemize}
\item Restrict tensor parallelism to a single node (i.e., $TP \leq 8$ on \SMNGPtwo).
\item Use enough micro-batches to keep the pipeline full and minimize bubble overhead.
\item Scale out primarily via data parallelism once the model parallel efficiency is saturated.
\end{itemize}

SuperMUC-NG Phase 2 operates in a power-capped production mode; in particular, Intel Data Center GPU Max~1550 accelerators run below nominal TDP.
\cite{intel_gpu_max_1550,power_capped_lrz_2025}
Accordingly, the reported performance reflects sustained production conditions rather than unconstrained peak operation.
A quantitative analysis of energy efficiency is left to future work.

\section*{Acknowledgments}
Ajay Navilarekal Rajgopal would like to thank his colleagues, Arjun Parab for help with the tooling used to parse the resulting log files, and Jophin John for guidance at various stages of this work.

\nocite{*}
\bibliographystyle{splncs04}
\bibliography{references} 

\end{document}